\documentclass[aps,prl,twocolumn,superscriptaddress,longbibliography]{revtex4-2}

\usepackage{amsmath,amssymb,amsfonts,amsthm}
\usepackage{color}
\usepackage[colorlinks,citecolor=blue,linkcolor=blue,urlcolor=blue]{hyperref}
\usepackage{verbatim}
\usepackage{graphicx}

\begin{document}

\title{Boundary Renormalization Group Flow of Entanglement Entropy at a (2+1)-Dimensional Quantum Critical Point}

\author{Zhiyan Wang}
\affiliation{State Key Laboratory of Surface Physics and Department of Physics, Fudan University, Shanghai 200438, China}
\affiliation{Department of Physics, School of Science and Research Center for Industries of the Future, Westlake University, Hangzhou 310030,  China}
\affiliation{Institute of Natural Sciences, Westlake Institute for Advanced Study, Hangzhou 310024, China}

\author{Zhe Wang}
\affiliation{Department of Physics, School of Science and Research Center for Industries of the Future, Westlake University, Hangzhou 310030,  China}
\affiliation{Institute of Natural Sciences, Westlake Institute for Advanced Study, Hangzhou 310024, China}

\author{Yi-Ming Ding}
\affiliation{State Key Laboratory of Surface Physics and Department of Physics, Fudan University, Shanghai 200438, China}
\affiliation{Department of Physics, School of Science and Research Center for Industries of the Future, Westlake University, Hangzhou 310030,  China}
\affiliation{Institute of Natural Sciences, Westlake Institute for Advanced Study, Hangzhou 310024, China}

\author{Zenan Liu}
\affiliation{Department of Physics, School of Science and Research Center for Industries of the Future, Westlake University, Hangzhou 310030,  China}
\affiliation{Institute of Natural Sciences, Westlake Institute for Advanced Study, Hangzhou 310024, China}

\author{Zheng Yan}
\email{zhengyan@westlake.edu.cn}
\affiliation{Department of Physics, School of Science and Research Center for Industries of the Future, Westlake University, Hangzhou 310030,  China}
\affiliation{Institute of Natural Sciences, Westlake Institute for Advanced Study, Hangzhou 310024, China}

\author{Long Zhang}
\email{longzhang@ucas.ac.cn}
\affiliation{Kavli Institute for Theoretical Sciences and CAS Center for Excellence in Topological Quantum Computation, University of Chinese Academy of Sciences, Beijing 100190, China}

\begin{abstract}
We investigate the second order R\'enyi entanglement entropy at the quantum critical point of spin-1/2 antiferromagnetic Heisenberg model on a columnar dimerized square lattice. The universal constant $\gamma$ in the area-law scaling $S_{2}(\ell) = \alpha\ell - \gamma$ is found to be sensitive to the entangling surface configurations, with $\gamma_{\text{sp}} > 0$ for strong-bond-cut (special) surfaces and $\gamma_{\text{ord}} < 0$ for weak-bond-cut (ordinary) surfaces, which is attributed to the distinct conformal boundary conditions. Introducing boundary dimerization drives a renormalization group (RG) flow from the special to the ordinary boundary criticality, and the constant $\gamma$ decreases monotonically with increasing dimerization strength, demonstrating irreversible evolution under the boundary RG flow. These results provide numerical evidence for a higher-dimensional analog of the $g$-theorem, and suggest $\gamma$ as a possible characteristic function for boundary RG flow in (2+1)-dimensional conformal field theory.
\end{abstract}

\date{\today}

\maketitle

\textit{\color{blue}Introduction.---}
The renormalization group (RG) provides a fundamental framework for understanding quantum field theory (QFT) and statistical physics. Under RG transformations, high-energy degrees of freedom (d.o.f.) are progressively integrated out, while effective interactions among the remaining low-energy d.o.f. are renormalized. This process generates a continuous RG flow in theory space, culminating in a conformal field theory (CFT) at the RG fixed point. The irreversibility of RG flow is heuristically motivated by the monotonic reduction of physical d.o.f. This intuition is rigorously formalized by a family of theorems initiated by Zamolodchikov's $c$-theorem for two-dimensional QFT~\cite{Zamolodchikov1986}, and subsequently extended to higher dimensions~\cite{Jafferis2011, Casini2012, Cardy1988, Komargodski2011, Casini2017}. These theorems establish that specific characteristic functions in theory space decrease monotonically along RG trajectories, imposing stringent nonperturbative constraints on possible RG flows and phase diagram structures.

Remarkably, properly regularized entanglement entropy (EE) serves as such a characteristic function~\cite{Casini2004a, Casini2012, Casini2017, Liu2013, Liu2014a, Nishioka2018}. For $(2+1)$-dimensional QFT, the relevant quantity is defined as $\mathcal{F}(R) = (R \partial_{R} - 1) S(R)$, where $S(R)$ denotes the EE of a smooth spatial subsystem of characteristic size $R$~\cite{Casini2012}. The EE exhibits the area-law scaling $S(R) = \alpha R - \gamma$ as $R \to \infty$, where $\alpha$ is a nonuniversal coefficient depending on the short-distance cutoff, whereas the constant $\gamma$ is universal and decreases monotonically under RG flow.

In quantum many-body systems with boundaries, the boundary interactions generally differ from the bulk interactions. When the bulk is tuned to a RG fixed point described by a CFT, the boundary interactions can undergo RG flows connecting distinct fixed points known as conformal boundary conditions, each of which corresponds to a universality class of boundary critical behavior~\cite{Cardy1984}. 

A fundamental question concerns the irreversibility of such boundary RG flows~\cite{Andrei2020}. In (1+1)-dimensional CFT, this irreversibility is governed by the $g$-theorem~\cite{Affleck1991, Friedan2004}, which establishes the boundary entropy $\ln g$~\cite{Affleck1991} as a monotonic function under RG flow. Crucially, $\ln g$ also has a quantum information interpretation via the EE~\cite{Casini2016}: For a spatial interval of length $r$ adjacent to the boundary, the EE takes the form $S(r) = (c/6) \ln (r/\epsilon) + c_{0} + \ln g$, where $c$ is the central charge of the bulk CFT, $\epsilon$ is a nonuniversal short-distance cutoff, and $c_{0}$ is a constant contribution from the bulk independent of the boundary condition. The $g$-theorem dictates that $\ln g$ decreases monotonically along the boundary RG flow. The irreversibility of boundary RG flow in (2+1)-dimensional CFT is established by examining the EE of a semicircular subsystem of radius $r$ adjacent to the boundary. The entangling surface intersects with the physical open boundary at two corners, contributing a universal logarithmic term proportional to $\ln(r/\epsilon)$ in the EE~\cite{Casini2007a}, whose coefficient serves as a monotonic characteristic function and decreases with the boundary RG flow~\cite{Jensen2016, Casini2019}.

In this work, we investigate the boundary RG flow at a $(2+1)$-dimensional quantum critical point (QCP) by monitoring the second R\'enyi EE of a \emph{cylinder} subsystem. The QCP is realized in a square-lattice columnar dimerized antiferromagnetic Heisenberg model, described by the 3D $\mathrm{O}(3)$ Wilson-Fisher CFT. Distinct conformal boundary conditions corresponding to the ordinary and the special boundary criticality are engineered by cutting weak or strong bonds at the boundary, respectively~\cite{Ding2018, weber2018nonordinary}. For a smooth entangling surface, the EE exhibits the area-law scaling~\cite{Eisert2010},
\begin{equation}
S(\ell) = \alpha \ell - \gamma, \quad \ell \to \infty,
\label{eq:area}
\end{equation}
where $\ell$ denotes the entangling surface length. We find that the universal constant $\gamma$ depends on the boundary condition at the entangling surface: $\gamma_{\text{sp}} > 0$ and $\gamma_{\text{ord}}<0$ for the special and the ordinary boundary conditions, respectively. Introducing dimerized interactions near the entangling surface triggers a boundary RG flow from the special to the ordinary criticality. Along this flow, $\gamma$ decreases monotonically. These results provide numerical evidence for a $(2+1)$-dimensional analog of the $g$-theorem, suggesting the universal term $\gamma$ in the EE as a candidate monotonic characteristic function under boundary RG flow.

\begin{figure}[!tb]
\centering
\includegraphics[width=\columnwidth]{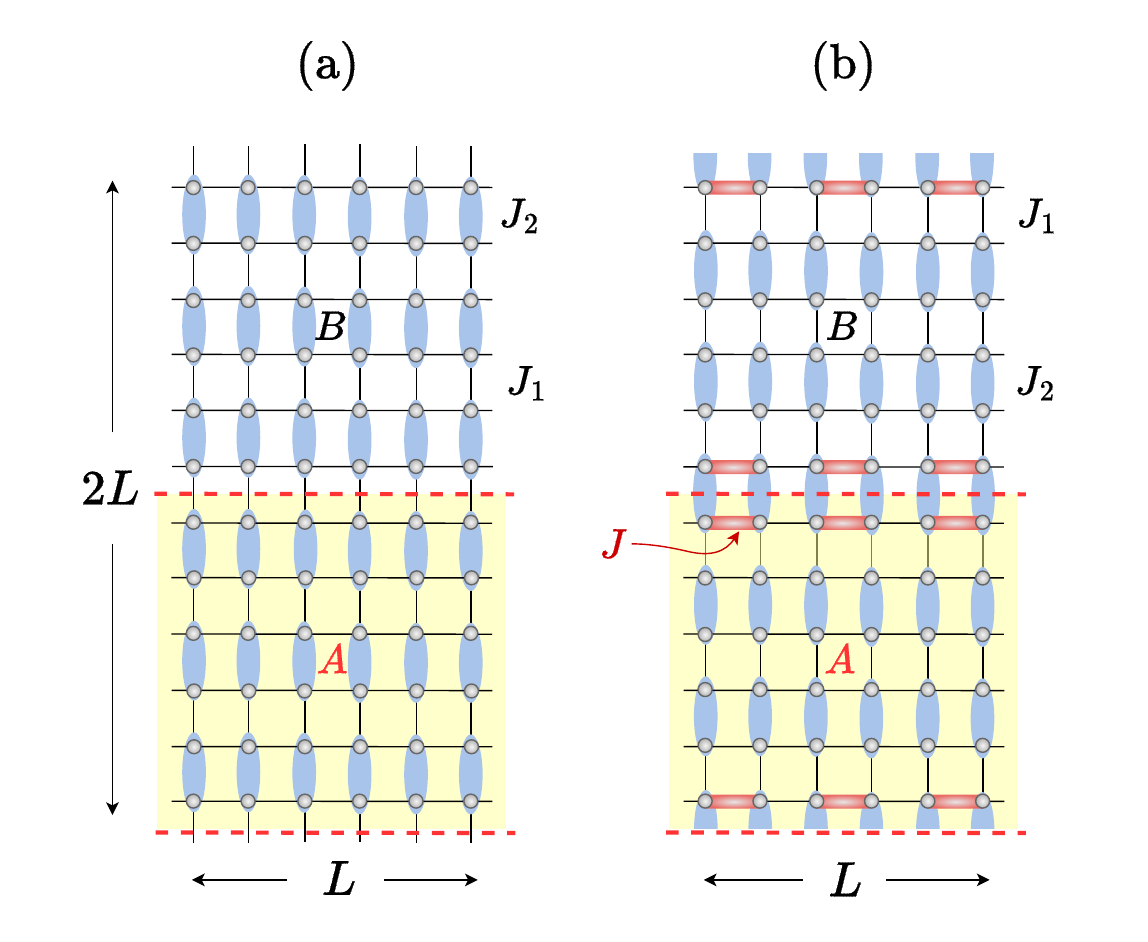}
\caption{Spin-$1/2$ antiferromagnetic Heisenberg model on a columnar dimerized square lattice. Thin black and thick blue bonds denote weak ($J_{1}$) and strong ($J_{2}$) exchange couplings, respectively. We fix $J_{2}/J_{1} = 1.9095$ at the QCP with $J_{1}=1$ setting the energy scale. The entangling surfaces of subsystem $A$ are formed by cutting weak bonds in (a) and cutting strong bonds in (b), corresponding to the ordinary and the special boundary conditions, respectively. The boundary dimerization in (b) is introduced on the thick red bonds with tunable coupling $J \geq 1$. When $J>1$, this explicitly gaps the dangling spin chains and drives a boundary RG flow from the special to the ordinary boundary condition.}
\label{fig:model}
\end{figure}

\textit{\color{blue} Model and method.---}
We investigate the spin-$1/2$ antiferromagnetic Heisenberg model on a square lattice with columnar dimerization illustrated in Fig.~\ref{fig:model}. The Hamiltonian is
\begin{equation}
H =  J_{1} \sum_{\langle ij\rangle} \mathbf{S}_{i} \cdot \mathbf{S}_{j} + J_{2} \sum_{\langle ij\rangle'} \mathbf{S}_{i} \cdot \mathbf{S}_{j},
\end{equation}
where $\langle ij \rangle$ and $\langle ij \rangle'$ denote nearest-neighbor spin pairs on the weak (thin black lines) and strong (thick blue lines) dimer bonds, respectively, as indicated in Fig.~\ref{fig:model}.

This system exhibits two distinct phases: For $J_{2}/J_{1} \gg 1$, strong dimerization results in a gapped disordered ground state; For $J_{2}/J_{1} \simeq 1$, long-range antiferromagnetic order emerges. These phases are separated by a QCP at $J_{2}/J_{1} = 1.90951(1)$~\cite{Ma2018}, which belongs to the 3D $\mathrm{O}(3)$ Wilson-Fisher universality class~\cite{Matsumoto2001a, Sandvik2010, Yasuda2013, Ma2018}. Throughout this work, we set $J_{1}=1$ as the energy scale and maintain $J_{2}$ at the critical value.

Although the bulk QCP is conventional, this model supports distinct boundary critical behaviors with different physical open boundary configurations~\cite{Ding2018}. As shown in Fig.~\ref{fig:model}(a), when the boundaries are created by cutting weak bonds, the system exhibits ordinary boundary criticality. In contrast, boundaries formed by cutting strong bonds shown in Fig.~\ref{fig:model}(b) yield special critical behavior, arising from the hybridization between the dangling spin chains and the bulk critical state~\cite{Zhang2017, Ding2018, weber2018nonordinary, Jian2021}.

To probe the boundary RG flow, we introduce explicit dimerized interactions along the strong-bond-cut boundaries. As depicted in Fig.~\ref{fig:model}(b), this is implemented through the thick red bonds parallel to the boundaries with tunable exchange coupling $J \geq 1$. For $J > 1$, the dimerized interactions gap out the dangling spin chains, driving a crossover from the special to the ordinary boundary criticality, which will be demonstrated later. Crucially, with the bulk fixed at the QCP, this setup enables direct investigation of the boundary RG flow purely induced by the boundary dimerization.  Our construction of introducing a dimerization pattern at the entangling boundary is consistent with the standard framework of boundary renormalization group (RG) theory~\cite{Affleck1991, Friedan2004}. Here, the dimerization $J$ acts as a tunable relevant boundary perturbation that drives a continuous flow from special to ordinary criticality while the bulk remains at the QCP. This approach aligns with established numerical paradigms where boundary physics is probed by engineering effective boundaries within the bulk.~\cite{Levin2006, Zhou2006PRA}

We compute the second R\'enyi EE defined as
\begin{equation}
S_{2} = -\ln \operatorname{tr} \rho_A^{2},
\end{equation}
where $\rho_A$ denotes the reduced density matrix of the subsystem $A$. Based on the stochastic series expansion (SSE) quantum Monte Carlo (QMC)~\cite{Sandvik1999, Sandvik2010,yan2019sweeping,yan2020improved}, we scan the EE in a large region of parameters via the bipartite reweight-annealing method \cite{Wang2025, Wang2024d,Ding2024,ding2025tracking} combined with the nonequilibrium work method~\cite{Alba2017a, DEmidio2020, Zhao2022b}. Simulations are performed with periodic boundary conditions on $L \times 2L$ square lattices. The subsystem $A$ is a $L \times L$ cylinder with entangling surfaces of total length $\ell = 2L$. To access ground state properties, we scale the inverse temperature linearly with system size, $\beta = 2L$.

\begin{figure}[!tb]
\centering
\includegraphics[width=\columnwidth]{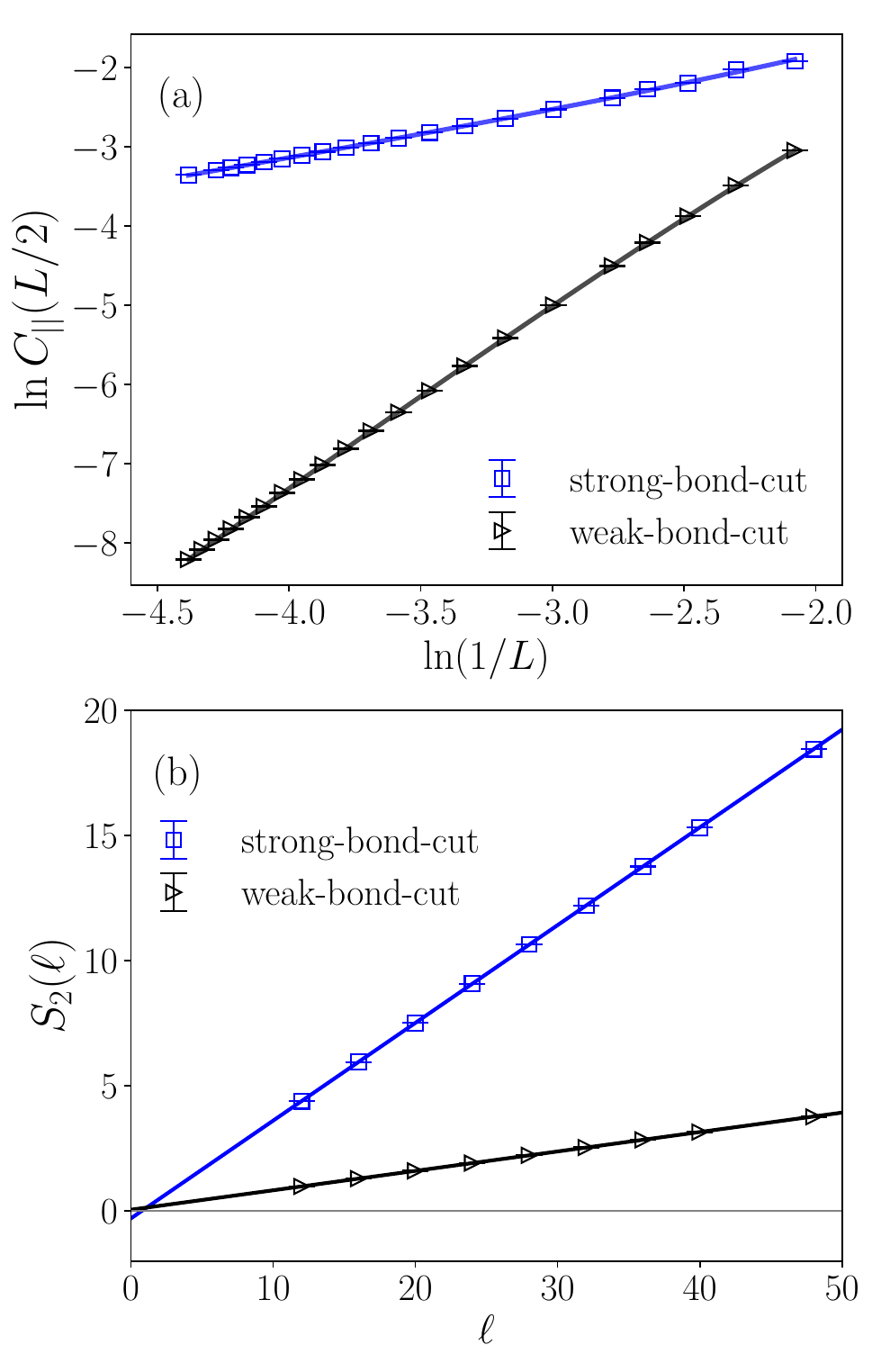}
\caption{Effects of distinct boundary conditions. (a) Spin correlations $C_{\parallel}(L/2)$ along physical open boundaries versus system size $L$ in log-log scales. Solid curves show power-law fitting with a subleading correction, $C_{\parallel}(L/2) = L^{-(1+\eta_{\parallel})}(b_{0}+b_{1}L^{-1})$. These data are reproduced from Ref.~\cite{Ding2018}. (b) Second R\'enyi entropy $S_{2}(\ell)$ versus entangling surface length $\ell$. Solid lines show the linear fitting to the area law $S(\ell) = \alpha\ell - \gamma$.}
\label{fig:rp}
\end{figure}

\textit{\color{blue} EE across distinct entangling surfaces.---}
We first examine how entangling surface configurations affect the entanglement structure. For QFT in flat spacetime, the entanglement Hamiltonian $H_{E} = -\ln \rho_{A}$ for a half-space corresponds to the Lorentz boost generator in that region~\cite{Casini2011, Cardy2016}. This implies a deep connection between the entanglement spectrum and the energy spectrum of a system with a physical open boundary. However, this mapping requires consistent regularization at the entangling surface, which is physically equivalent to imposing a specific boundary condition~\cite{Ohmori2015, Cardy2016}. While lattice models provide a natural regularization through their intrinsic short-distance cutoff, the emergent boundary condition depends sensitively on both the lattice Hamiltonian and the entangling surface configuration. In (1+1)-dimensional CFTs, the conformal boundary condition at an entangling surface typically matches that at a physical open boundary~\cite{Lauchli2013, Yu2022}. We extend this conjecture to our (2+1)-dimensional system, proposing that the weak-bond-cut entangling surfaces yield ordinary boundary condition, the strong-bond-cut boundaries yield special boundary condition~\cite{zhu2025bipartite,zyan2021entanglement,liu2023probing,song2023different}, and the dimerization on the strong-bond-cut boundaries drives a crossover from the special to the ordinary boundary condition as the dimerization strength increases.

In the presence of a physical open boundary, the boundary spin correlation function follows a power-law decay: $C_{\parallel}(r) = \langle \mathbf{S}_{x+r} \cdot \mathbf{S}_{x} \rangle \propto (-1)^r r^{-(1+\eta_{\parallel})}$, where the anomalous dimension $\eta_{\parallel}$ characterizes the universality class of boundary critical behavior~\cite{Binder1983phase}. As established in Ref.~\cite{Ding2018} and reproduced in Fig.~\ref{fig:rp}(a), our system exhibits two distinct boundary universality classes: Weak-bond-cut boundaries show ordinary criticality with $\eta_{\parallel} = 1.387(4)$; In contrast, strong-bond-cut boundaries show nonordinary criticality with $\eta_{\parallel} = -0.445(15)$, which is consistent with the special transition~\cite{Diehl1981, Diehl1986phase}.

Figure~\ref{fig:rp}(b) presents the second R\'enyi entropy $S_{2}(\ell)$ as a function of the entangling surface length $\ell$ for both weak- and strong-bond-cut entangling surface configurations. In both cases, $S_{2}(\ell)$ increases linearly with $\ell$, confirming the area-law behavior. Fitting to Eq.~(\ref{eq:area}) yields distinct coefficients: $\alpha_{\text{sp}} = 0.3906(1)$ and $\gamma_{\text{sp}} = 0.292(4)$ for strong-bond-cut surfaces, whereas $\alpha_{\text{ord}} = 0.0776(1)$ and $\gamma_{\text{ord}} = -0.056(3)$ for weak-bond-cut surfaces. We observe that $S_{2}(\ell)$ is significantly enhanced for strong-bond-cut surfaces, which is primarily driven by the linear term $\alpha_{\text{sp}} > \alpha_{\text{ord}}$. The linear term originates mainly from short-range correlations across the entangling surfaces and is inherently nonuniversal. In contrast, the constant $\gamma$ is independent of short-distance regularization and subsystem geometry for smooth boundaries, thus is nominally universal and expected to characterize the bulk QCP~\cite{Metlitski2009}. Remarkably, we observe an unexpected sign reversal of the constant: $\gamma_{\text{sp}} > 0$ and $\gamma_{\text{ord}}<0$. This difference emerges despite the identical bulk critical state. Therefore, we attribute the discrepancy in $\gamma$ to the distinct conformal boundary conditions realized at the entangling surfaces.

\begin{figure*}[!tb]
\centering
\includegraphics[width=\textwidth]{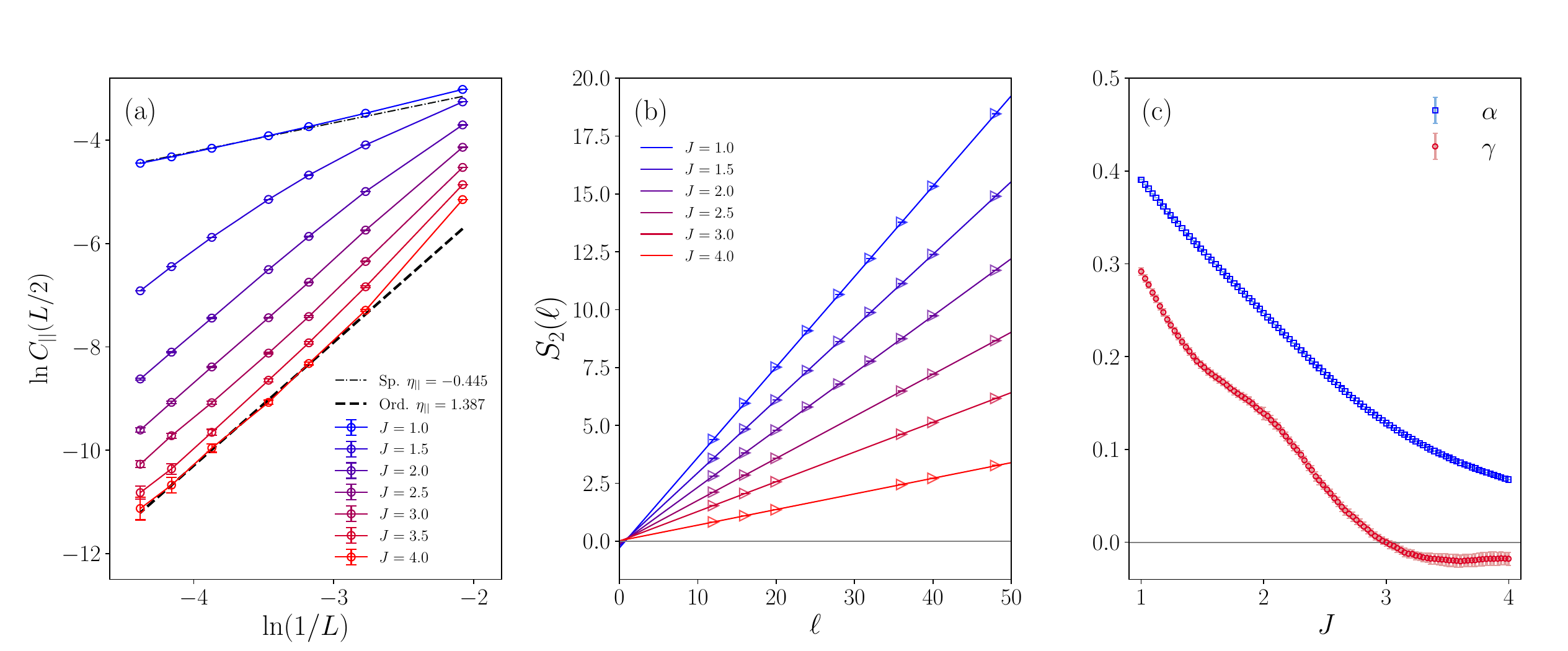}
\caption{Effects of boundary RG flow driven by dimerization strength $J$ at strong-bond-cut surfaces. (a) Boundary spin correlations $C_{\parallel}(L/2)$ versus system size $L$ in log-log scales. The power-law scaling of the special and the ordinary boundary critical behaviors are plotted as the dotted and the dashed lines for comparison. (b) Second R\'enyi EE $S_{2}(\ell)$ versus entangling surface length $\ell$ for varying $J$. (c) Extracted coefficients from the area-law fitting versus dimerization strength $J$. The nonuniversal coefficient $\alpha$ decreases with increasing $J$, while the universal constant $\gamma$ also decreases monotonically, saturating for $J \gtrsim 3.5$ within error bars.}
\label{fig:entropy}
\end{figure*}

\textit{\color{blue}Boundary RG flow.---}
We now investigate how the explicit dimerization along strong-bond-cut boundaries drives the boundary RG flow. Introducing dimerization acts as a relevant perturbation that immediately gaps the dangling spin chains. This gap formation fundamentally alters the boundary critical behavior. While the special criticality requires coupling between the gapless dangling chain and bulk critical modes, a gapped chain coupled to the bulk can only support ordinary boundary condition~\cite{Jian2021, Song2025}. Consequently, dimerization triggers a crossover from the special to the ordinary boundary criticality. This crossover is quantified through the boundary spin correlations $C_{\parallel}(L/2)$ plotted in Fig.~\ref{fig:entropy}(a). 
As the dimerization strength $J$ increases, the slope of $C_{\parallel}(L/2)$ in log-log scales evolves continuously for the available system sizes. It is expected that for any fixed $J > 1$, the boundary spin correlations will asymptotically flow towards the ordinary fixed point characterized by $\eta_{||} = 1.387$ in the large-$L$ limit.  This demonstrates that the boundary dimerization serves as a relevant perturbation, driving an RG flow from the special to the ordinary criticality.

The second R\'enyi entropy $S_{2}(\ell)$ under varying boundary dimerization strengths is presented in Fig.~\ref{fig:entropy}(b). Crucially, $S_{2}(\ell)$ maintains area-law scaling for all $J$ values. Through linear fitting to Eq.~(\ref{eq:area}), we extract the coefficients $\alpha$ and $\gamma$ at each dimerization strength and plot their evolution in Fig.~\ref{fig:rp}(c).

We observe systematic suppression of $S_{2}(\ell)$ with increasing dimerization, which can be attributed to the enhanced intra-dimer entanglement at the boundary. This reduces the quantum correlations across the entangling surface, which manifests as a monotonic decrease in the nonuniversal coefficient $\alpha$ with increasing $J$.

Remarkably, the universal constant $\gamma$ also exhibits a monotonic decrease with increasing $J$, saturating within error bars in the strong dimerization limit ($J \gtrsim 3.5$). While $\gamma$ remains independent of short-distance regularization and subsystem geometry, its evolution directly tracks the boundary RG flow driven by dimerization. 
The observed monotonic decrease of $\gamma$ suggests its role as a characteristic function for boundary RG flow in (2+1) dimensions, analogous to the boundary entropy $\ln g$ in (1+1)-dimensional CFTs~\cite{Affleck1991, Friedan2004, Casini2016} and suggests a higher-dimensional anolog of $g$-theorem. This is consistent with the irreversibility of boundary RG flows; However, we note that $\gamma$ is different from the $b$ coefficient in the entropic $b$-theorem established in Ref.~\cite{Casini2019}, which is defined by the boundary Weyl anomaly coefficient and can be extracted from the logarithmic term in the von Neumann EE of a semicircular region close to the boundary.

\textit{\color{blue} Conclusion and discussions.---}
We have investigated the second R\'enyi EE $S_{2}(\ell)$ at the QCP of a columnar dimerized antiferromagnetic Heisenberg model on the square lattice. Our central discovery reveals that the universal constant term $\gamma$ in the area-law scaling $S_{2}(\ell) = \alpha\ell - \gamma$ exhibits remarkable sensitivity to the entangling surface configurations. Specifically, we observe that $\gamma_{\text{sp}} > 0$  for the strong-bond-cut entangling surfaces, and $\gamma_{\text{ord}} < 0$ for the weak-bond-cut surfaces. Through the correspondence between the entangling surfaces and the physical open boundaries, we attribute this difference to the distinct conformal boundary conditions. Moreover, introducing dimerization along the strong-bond-cut surfaces drives a controlled boundary RG flow from the special to the ordinary criticality. Along this flow, $\gamma$ decreases monotonically with the dimerization strength $J$, saturating in the strong dimerization regime. Therefore, we propose that the boundary RG flow at a (2+1)-dimensional QCP exhibits irreversible behavior, and the constant $\gamma$ serves as a monotonic characteristic function of the boundary RG flow. Our work thus provides numerical evidence for a higher-dimensional analog of the $g$-theorem, and identify the universal constant of the EE as the characteristic function directly analogous to the role of boundary entropy in (1+1)-dimensional CFT.

Our findings raise several open questions deserving further investigation.

First, the interpretation of $\gamma$ as a universal constant independent of short-distance regularization and subsystem geometry rests on field-theoretic arguments\cite{Metlitski2009}. However, a systematic verification is essential through comparative studies with distinct lattice model realizations of the three-dimensional O(3) universality class and tests on the geometric dependence using anisotropic subsystems with different aspect ratios. Such examination is expected to establish $\gamma$ as a genuine universal quantity characterizing the bulk CFT and the conformal boundary condition.

Second, while our spin correlation data demonstrate the evolution of the physical open boundary from the special to the ordinary criticality, the observed $\gamma$ saturation value under strong dimerization still deviates from the $\gamma_{\text{ord}}$ value obtained with weak-bond-cut entangling surfaces. This discrepancy might be attributed to the finite-size effect, which is particularly significant for strong dimerization. High-precision R\'enyi EE data at larger system sizes, particularly in the strong dimerization regime are required to clarify this discrepancy. The extracted values of $\gamma$ are subject to finite-size corrections, which are expected to scale as $1/L$ or $1/\ell$. Despite these corrections, our results demonstrate a robust monotonic evolution and a clear sign change of $\gamma$. We argue that $\gamma$ does not vanish for $J > 0$ in the thermodynamic limit; instead, it remains a finite nonzero value characterizing the specific conformal boundary condition, as evidenced by the distinct values $\gamma_{sp} = 0.292(4)$ and $\gamma_{ord} = -0.056(3)$. This supports the role of $\gamma$ as a meaningful characteristic function for tracking boundary RG flows.

Third, our use of the second R\'enyi EE $S_{2}$ is motivated by its computational accessibility in QMC simulations of large lattices. While the von Neumann EE $S = -\operatorname{tr}(\rho_{A}\ln\rho_{A})$ holds deeper theoretical significance, particularly in generalized $c$-theorems and $g$-theorems because of its strong subadditivity (SSA) inequality~\cite{Casini2004a, Casini2012, Casini2017, Casini2016, Casini2019}, both EEs exhibit the area-law scaling with universal constant terms at QCPs. Moreover, both EEs are derived from the entanglement spectrum, and the replica trick establishes its relation with the $n$th R\'enyi EE, $S = \lim_{n\to 1} S_{n}$, suggesting that the observed entangling surface dependence and the monotonic decrease under boundary RG flow of the constant term in $S_{2}$ may extend to the von Neumann EE.  Our findings thus motivate further research into the von Neumann counterpart. Although direct computation of $S$ at (2+1)-dimensional QCPs remains challenging for large systems, emerging tensor network methods offer promising pathways for future verification. 

Fourth, our proposal of $\gamma$ as a boundary RG characteristic function can be generalized to other universality classes of (2+1)-dimensional QCPs and diverse boundary deformation protocols. Ultimately, a rigorous proof on the monotonic decrease of the constant $\gamma$ under boundary RG flow for (2+1)-dimensional CFT deserves further theoretical investigation.

\textit{\color{blue} Acknowledgements.---}
L.Z. is grateful to discussions with Xinan Zhou. We thank the helpful discussions with Wei Zhu and Yijian Zou.
L.Z. is supported by the National Natural Science Foundation of China (No. 12174387), the Chinese Academy of Sciences (Nos. YSBR-057 and JZHKYPT-2021-08), and the Innovative Program for Quantum Science and Technology (No. 2021ZD0302600). Zhe Wang is supported by the China Postdoctoral Science Foundation under Grants No.2024M752898. Z.L. is supported by the China Postdoctoral Science
Foundation under Grants No.2024M762935 and NSFC Special Fund for Theoretical Physics under Grants No.12447119.
The work is supported by the Scientific Research Project (No. WU2025B011) and the Start-up Funding of Westlake University. 
The authors thank the high-performance computing center of Westlake University and the Beijng PARATERA Tech Co.,Ltd. for providing HPC resources.

\textit{Data availability.} The data that support the findings of this article are openly available~\cite{zhiyan_2026_19014231}.

\bibliography{main}
\end{document}